\documentclass[proceedings]{JHEP3}
\usepackage{amsfonts}
\usepackage{amsmath}
\usepackage{epsfig,multicol}

\newbox\mybox

\newcommand\fverb{\setbox\mybox=\hbox\bgroup\verb}
\newcommand\fverbdo{\egroup\medskip\noindent\fbox{\unhbox\mybox}\ }
\newcommand\fverbit{\egroup\item[\fbox{\unhbox\mybox}]}
\conference{2D NC harmonic oscillator  }
\abstract{The two dimensional set of canonical relations giving rise to minimal uncertainties previously constructed from a
q-deformed oscillator algebra is further investigated. We provide a representation for this algebra in terms of a
flat noncommutative space and employ it to study the eigenvalue spectrum for the harmonic oscillator on this space. The
perturbative expression for the eigenenergy indicates that the model might possess an exceptional point at which the
spectrum becomes complex and its PT-symmetry is spontaneously broken.}

\title{The two dimensional harmonic oscillator on a noncommutative space
with minimal uncertainties}
\author{Sanjib Dey and Andreas Fring \\
Centre for Mathematical Science, City University London,\\
$\,\,$ Northampton Square, London EC1V 0HB, UK\\
E-mail: sanjib.dey.1@city.ac.uk, a.fring@city.ac.uk}

\input{tcilatex}
\begin{document}

In \cite{DFG} we demonstrated how canonical relations implying minimal
uncertainties can be derived from a $q$-deformed oscillator algebra for the
creation and annihilation operators $A_{i}^{\dagger }$, $A_{i}$ 
\begin{equation}
A_{i}A_{j}^{\dagger }-q^{2\delta _{ij}}A_{j}^{\dagger }A_{i}=\delta
_{ij},\quad \lbrack A_{i}^{\dagger },A_{j}^{\dagger }]=0,\quad \lbrack
A_{i},A_{j}]=0,\qquad \text{for }i,j=1,2,3;q\in \mathbb{R},  \label{AAAA}
\end{equation}%
as investigated for instance in \cite{Bieden,MacF,ChangPu,AFBB,AFLGBB}.
Starting from the general Ansatz%
\begin{eqnarray}
X &=&\hat{\kappa}_{1}(A_{1}^{\dagger }+A_{1})+\hat{\kappa}%
_{2}(A_{2}^{\dagger }+A_{2})+\hat{\kappa}_{3}(A_{3}^{\dagger }+A_{3}),
\label{an1} \\
Y &=&i\hat{\kappa}_{4}(A_{1}^{\dagger }-A_{1})+i\hat{\kappa}%
_{5}(A_{2}^{\dagger }-A_{2})+i\hat{\kappa}_{6}(A_{3}^{\dagger }-A_{3}), \\
Z &=&\hat{\kappa}_{7}(A_{1}^{\dagger }+A_{1})+\hat{\kappa}%
_{8}(A_{2}^{\dagger }+A_{2})+\hat{\kappa}_{9}(A_{3}^{\dagger }+A_{3}), \\
P_{x} &=&i\check{\kappa}_{10}(A_{1}^{\dagger }-A_{1})+i\check{\kappa}%
_{11}(A_{2}^{\dagger }-A_{2})+i\check{\kappa}_{12}(A_{3}^{\dagger }-A_{3}),
\\
P_{y} &=&\check{\kappa}_{13}(A_{1}^{\dagger }+A_{1})+\check{\kappa}%
_{14}(A_{2}^{\dagger }+A_{2})+\check{\kappa}_{15}(A_{3}^{\dagger }+A_{3}), \\
P_{z} &=&i\check{\kappa}_{16}(A_{1}^{\dagger }-A_{1})+i\check{\kappa}%
_{17}(A_{2}^{\dagger }-A_{2})+i\check{\kappa}_{18}(A_{3}^{\dagger }-A_{3}),
\label{an6}
\end{eqnarray}%
with $\hat{\kappa}_{i}=\kappa _{i}\sqrt{\hbar /(m\omega )}$ for $i=1,\ldots
,9$ and $\check{\kappa}_{i}=\kappa _{i}\sqrt{m\omega \hbar }$ for $%
i=10,\ldots ,18$ we constructed some particular solutions and investigated
the harmonic oscillator on these spaces. Here we provide an additional two
dimensional solution previously reported in \cite{AFLGBB}. Setting $\kappa
_{3}=$ $\kappa _{6}=$ $\kappa _{7}=\kappa _{12}=$ $\kappa _{15}=$ $\kappa
_{16}=$ $\kappa _{17}=$ $\kappa _{18}=0$ in equations (\ref{an1})-(\ref{an6}%
), employing the constraints reported in \cite{AFLGBB} together with the
subsequent nontrivial limit $q\rightarrow 1$, the deformed oscillator
algebra 
\begin{equation}
\begin{array}{lll}
\lbrack X,Y]=i\theta \left( 1+\hat{\tau}Y^{2}\right) ,\qquad & 
[X,P_{x}]=i\hbar \left( 1+\hat{\tau}Y^{2}\right) ,\qquad & [X,P_{y}]=0, \\ 
\lbrack P_{x},P_{y}]=i\hat{\tau}\frac{\hbar ^{2}}{\theta }Y^{2},~~ & 
[Y,P_{y}]=i\hbar \left( 1+\hat{\tau}Y^{2}\right) ,~\  & [Y,P_{x}]=0,%
\end{array}
\label{bbb}
\end{equation}%
was obtained, with $\hat{\tau}=\tau m\omega /\hbar $ having the dimension of
an inverse squared length. By the same reasoning as provided in \cite%
{Kempf1,Kempf2,AFBB,AFLGFGS,AFLGBB,DFG}, we find the minimal uncertainties%
\begin{equation}
\Delta X_{\min }=\left\vert \theta \right\vert \sqrt{\hat{\tau}+\hat{\tau}%
^{2}\left\langle Y\right\rangle _{\rho }^{2}},~~\Delta Y_{\min }=0,~~\Delta
\left( P_{x}\right) _{\min }=0,~~\Delta \left( P_{y}\right) _{\min }=\hbar 
\sqrt{\hat{\tau}+\hat{\tau}^{2}\left\langle Y\right\rangle _{\rho }^{2}},
\end{equation}%
where $\left\langle .\right\rangle _{\rho }$ denotes the inner product on a
Hilbert space with metric $\rho $ in which the operators $X,Y,P_{x}$ and $%
P_{y}$ are Hermitian. So far no representation for the two dimensional
algebra (\ref{bbb}) was provided. Here we find that it can be represented by%
\begin{equation}
X=x_{0}+\hat{\tau}y_{0}^{2}x_{0},\quad Y=y_{0},\quad P_{x}=p_{x_{0}},\quad 
\text{and\quad }P_{y}=p_{y_{0}}-\hat{\tau}\frac{\hbar }{\theta }%
y_{0}^{2}x_{0},  \label{rep2}
\end{equation}%
where the $x_{0},y_{0},p_{x_{0}},p_{y_{0}}$ satisfy the common commutation
relations for the flat noncommutative space%
\begin{equation}
\begin{array}{lll}
\lbrack x_{0},y_{0}]=i\theta , & [x_{0},p_{x_{0}}]=i\hbar ,\qquad \quad & 
[x_{0},p_{y_{0}}]=0, \\ 
\lbrack p_{x_{0}},p_{y_{0}}]=0,\qquad \quad & [y_{0},p_{y_{0}}]=i\hbar , & 
[y_{0},p_{x_{0}}]=0,%
\end{array}%
~~\ \ \ \ \ \text{for }\theta \in \mathbb{R}.  \label{cano}
\end{equation}

Clearly there exist many more solutions one may construct in this systematic
manner from the Ansatz (\ref{an1})-(\ref{an6}), which will not be our
concern here. Instead we will study a concrete model, i.e. the
two-dimensional harmonic oscillator on the noncommutative space described by
the algebra (\ref{bbb}). Using the representation (\ref{rep2}), the
corresponding Hamiltonian reads 
\begin{eqnarray}
H_{ncho}^{2D} &=&\frac{1}{2m}(P_{x}^{2}+P_{y}^{2})+\frac{m\omega ^{2}}{2}%
(X^{2}+Y^{2})  \label{2dho} \\
&=&H_{fncho}^{2D}+\frac{\hat{\tau}}{2}\left[ m\omega
^{2}\{y_{0}^{2}x_{0},x_{0}\}-\frac{\hbar }{m\theta }%
\{y_{0}^{2}x_{0},p_{y_{0}}\}\right] +\frac{\hat{\tau}^{2}}{2}\left[ m\omega
^{2}+\frac{\hbar ^{2}}{m\theta ^{2}}\right] y_{0}^{2}x_{0}y_{0}^{2}x_{0} 
\notag
\end{eqnarray}%
where we used the standard notation for the anti-commutator $\left\{
A,B\right\} :=AB+BA$. Evidently this Hamiltonian is non-Hermitian with
regard to the standard inner product, but respects an antilinear symmetry $%
\mathcal{PT}_{\pm }$: $x_{0}\rightarrow \pm x_{0}$, $y_{0}\rightarrow \mp
y_{0}$, $p_{x_{0}}\rightarrow \mp p_{x_{0}}$, $p_{y_{0}}\rightarrow \pm
p_{y_{0}}$ and $i\rightarrow -i$. This suggests that its eigenvalue spectrum
might be real, or at least real in parts \cite{Bender:1998ke,AliI,Benderrev}%
. Let us now investigate the spectrum perturbatively around the solution of
the standard harmonic oscillator. In order to perform such a computation we
need to convert flat noncommutative space into the standard canonical
variable $x_{s}$, $y_{s}$, $p_{x_{s}}$ and $p_{y_{s}}$. This is achieved by
means of a so-called Bopp-shift $x_{0}\rightarrow x_{s}-\frac{\theta }{\hbar 
}p_{y_{s}}$, $y_{0}\rightarrow y_{s}$,\ $p_{x_{0}}\rightarrow p_{x_{s}}$ and 
$p_{y_{0}}\rightarrow p_{y_{s}}$. The Hamiltonian in (\ref{2dho}) then
acquires the form%
\begin{eqnarray}
H_{ncho}^{2D} &=&H_{ho}^{2D}+\frac{m\theta ^{2}\omega ^{2}}{2\hbar ^{2}}%
p_{y_{s}}^{2}-\frac{m\theta \omega ^{2}}{2\hbar }\{x_{s},p_{y_{s}}\}+\frac{%
\hat{\tau}}{2}\left[ m\omega ^{2}\{y_{s}^{2}x_{s},x_{s}\}-\frac{\hbar }{%
m\theta }\{y_{s}^{2}x_{s},p_{y_{s}}\}\right]  \\
&&+\frac{\hat{\tau}}{2}\left[ \left( \frac{1}{m}+\frac{m\theta ^{2}\omega
^{2}}{\hbar ^{2}}\right) \{y_{s}^{2}p_{y_{s}},p_{y_{s}}\}-\frac{m\theta
\omega ^{2}}{\hbar }\left(
\{y_{s}^{2}p_{y_{s}},x_{s}\}+\{y_{s}^{2}x_{s},p_{y_{s}}\}\right) \right] ~~ 
\notag \\
&&-\frac{\hat{\tau}^{2}}{2}\left[ \frac{m\theta \omega ^{2}}{\hbar }+\frac{%
\hbar }{m\theta }\right] \left(
y_{s}^{2}p_{y_{s}}y_{s}^{2}x_{s}+y_{s}^{2}x_{s}y_{s}^{2}p_{y_{s}}\right) +%
\frac{\hat{\tau}^{2}}{2}\left[ \frac{1}{m}+\frac{m\theta ^{2}\omega ^{2}}{%
\hbar ^{2}}\right] y_{s}^{2}p_{y_{s}}y_{s}^{2}p_{y_{s}}~~  \notag \\
&&+\frac{\hat{\tau}^{2}}{2}\left[ m\omega ^{2}+\frac{\hbar ^{2}}{m\theta ^{2}%
}\right] y_{s}^{2}x_{s}y_{s}^{2}x_{s}  \notag \\
&=&H_{ho}^{2D}(x_{s},y_{s},p_{x_{s}},p_{y_{s}})+H_{nc}^{2D}(x_{s},y_{s},p_{x_{s}},p_{y_{s}}).
\notag
\end{eqnarray}

In this formulation we may now proceed to expand perturbatively around the
standard two dimensional Fock space harmonic oscillator solution with
normalized eigenstates%
\begin{eqnarray}
\left\vert n_{1}n_{2}\right\rangle &=&\frac{(a_{1}^{\dagger
})^{n_{1}}(a_{2}^{\dagger })^{n_{2}}}{\sqrt{n_{1}!n_{2}!}}\left\vert
00\right\rangle ,\quad a_{i}^{\dagger }\left\vert n_{1}n_{2}\right\rangle =%
\sqrt{n_{i}+1}\left\vert (n_{1}+\delta _{i1})(n_{2}+\delta
_{i2})\right\rangle ,\quad \\
a_{i}\left\vert 00\right\rangle &=&0,~~~~~\qquad \qquad \ \ \ \ \ \ \ \
a_{i}\left\vert n_{1}n_{2}\right\rangle =\sqrt{n_{i}}\left\vert
(n_{1}-\delta _{i1})(n_{2}-\delta _{i2})\right\rangle ,
\end{eqnarray}%
for $i=1,2$, such that $H_{ho}^{2D}\left\vert nl\right\rangle
=E_{nl}^{(0)}\left\vert nl\right\rangle $. The energy eigenvalues for the
Hamiltonian $H_{ncho}^{2D}$ then result to 
\begin{eqnarray}
E_{nl}^{(p)} &=&E_{nl}^{(0)}+E_{nl}^{(1)}+E_{nl}^{(2)}+\mathcal{O}(\tau ^{2})
\\
&=&E_{nl}^{(0)}+\left\langle nl\right\vert H_{nc}^{2D}\left\vert
nl\right\rangle +\sum_{p,q\neq n+l=p+q}\frac{\left\langle nl\right\vert
H_{nc}^{2D}\left\vert pq\right\rangle \left\langle pq\right\vert
H_{nc}^{2D}\left\vert nl\right\rangle }{E_{nl}^{(0)}-E_{pq}^{(0)}}+\mathcal{O%
}(\tau ^{2})  \notag \\
&=&\omega \hbar \left( n+l+1\right) +\frac{1}{16}\hbar \omega \Omega \left[
2n-(2l+1)\Omega +10l+6\right]  \notag \\
&&+\frac{1}{8}\hbar \tau \omega \left[ \Omega \left(
8nl+4n+6l^{2}+10l+5\right) +10nl+5n+5l^{2}+10l+5\right] +\mathcal{O}(\tau
^{2}),  \notag
\end{eqnarray}%
where $\Omega =m^{2}\theta ^{2}\omega ^{2}/\hbar ^{2}$. We notice the minus
sign in one of the terms, which might be an indication for the existence of
an exceptional point \cite{CarlExPoint,Kato} in the spectrum. Naturally it
would be very interesting to obtain a more precise expression for the
eigenenergies, but nonetheless the first order approximations will be very
useful for the computation of coherent states \cite{SDAFprep}.


\noindent \textbf{Acknowledgments:} SD is supported by a City University
Research Fellowship.



\end{document}